\documentstyle[12pt,epsf]{article}
\def\be{\begin{equation}}
\def\ee{\end{equation}}
\def\beq{\begin{eqnarray}}
\def\eeq{\end{eqnarray}}
\def\e{\epsilon}
\def\l{\lambda}

\def\R{{\cal R}}
\def\RR{{\bf R^3}}
\def\({\left (}
\def\){\right )}
\def\[{\left [}
\def\[{\right ]}
\begin{document}

\begin{titlepage}
\bigskip
\rightline{}
\rightline{hep-th/0304199}
\bigskip\bigskip\bigskip\bigskip
\centerline{\Large \bf {Negative Energy Density in Calabi-Yau
Compactifications}}
     \bigskip\bigskip
      \bigskip\bigskip

  \centerline{\large Thomas Hertog${}^1$, Gary T. Horowitz${}^1$,
  and Kengo Maeda${}^2$}
     \bigskip\bigskip
  \centerline{\em ${}^1$ Department of Physics, UCSB, Santa Barbara, CA 93106}
  \centerline{hertog@vulcan.physics.ucsb.edu, 
   gary@physics.ucsb.edu }
	    \bigskip
 \centerline{\em ${}^2$ Yukawa Institute for Theoretical Physics,
  Kyoto University, Kyoto 606-8502, Japan}	    
	    \centerline{kmaeda@yukawa.kyoto-u.ac.jp}
	    \bigskip\bigskip

\begin{abstract}

We show that a large class of supersymmetric compactifications, including all
simply connected Calabi-Yau and $G_2$
manifolds, have classical configurations with negative energy density as
seen from four dimensions. In fact, the energy density can be
arbitrarily negative -- it is unbounded from below. Nevertheless, 
positive energy theorems show that the
total ADM energy remains positive. Physical consequences of the
negative 
energy density include new thermal instabilities, and possible violations
of cosmic censorship.

\end{abstract}
\end{titlepage}

\baselineskip=18pt

\setcounter{equation}{0}
 \section{Introduction}
 
Standard supersymmetric compactifications of string theory
consist of solutions of the form $M_4\times K$ where $M_4$ is
four dimensional Minkowski spacetime and $K$ is a compact, Ricci flat
manifold admitting a covariantly constant spinor. Familiar examples of $K$
include $T^n$, $K3$, Calabi-Yau spaces, and manifolds with $G_2$ holonomy.
It turns out that there is an important
qualitative difference between some of these supersymmetric
compactifications and others. We will show that a large class, 
including all simply connected
Calabi-Yau and $G_2$ manifolds, have  the surprising property
that there are configurations with negative energy
density. In other words, from a four dimensional perspective, there can
be finite regions of space with negative energy. In fact, the energy
density is unbounded from below! In contrast, these
properties do not hold - at least not in the same way - 
for $T^n$ or $K3$ compactifications.

We will work in the context of classical vacuum solutions to higher
dimensional general relativity.  So there is no matter or energy density in
the higher dimensional space. The negative energy arises from the Kaluza-Klein
compactification. These results clearly apply to string theory since
vacuum solutions are all (approximate) solutions to string
theory and M theory with the 
other fields set to zero. We should emphasize that the negative energy
regions we have in mind are very different from previous discussions
of negative energy in Kaluza-Klein theory \cite{Witten:gj,Brill:di}. 
In those cases,
supersymmetry was broken and the spacetime was not topologically a product
${\bf R}^4\times K$. In contrast, the
compactifications we consider here are all supersymmetric and the spacetime
will be a product manifold.

The key mathematical fact which allows configurations of negative energy density
is the following. As we review in section 2, all simply connected
compact manifolds of dimension five, six, or seven admit 
Riemannian metrics with
positive scalar curvature. Other manifolds, such as $T^n$ and $K3$ do not.
Positive scalar curvature on $K$ leads to negative energy density as follows.
Vacuum solutions can
be characterized by their initial data on $\RR \times K$.
 Since we want to minimize the energy, we
set the time derivatives of the metric to zero. For time symmetric
initial data, the Einstein constraint equations reduce to the vanishing
of the scalar curvature, $\R=0$.
For a product metric on $\RR \times K$, $\R = \R_3 + \R_K$ where $\R_3$ is
the scalar curvature on $\RR$ and $\R_K$ is the scalar curvature on $K$.
If $\R_K >0$, we must take $\R_3 <0$. But negative
scalar curvature on $\RR$ is just like negative energy density. (Recall that
the usual constraint of $3+1$ dimensional general relativity says\footnote{We
set $8\pi G =1$.}
$\R_3 = 2\rho$
in the time symmetric case.) Therefore, from an effective four dimensional
standpoint, positive scalar curvature on $K$ acts 
like negative energy density. In other words, ten dimensional
vacuum gravity has configurations with effective 
negative energy density! Of course, we must require that the metric on $K$ 
approaches the standard Ricci flat metric at infinity, so we cannot
keep the metric a product everywhere.  However, one can satisfy this
boundary condition and keep the region of negative energy density by
taking the  metric to be product inside
a large ball of radius $R_0$. In a finite transition region, $R_0 < r < R_1$,
one can change the metric on $K$ to the standard Ricci flat metric.

Not only is there negative energy density in four dimensions, but this
energy can be arbitrarily negative. This follows immediately from the
fact that there is no upper bound on the scalar curvature $\R_K$.
Given a  metric on $K$ with $\R_K>0$, one can clearly rescale it by
a constant factor and make the scalar curvature arbitrarily large.
This shows that the negative energy density is unbounded from below.
(Of course, once the curvature becomes larger than the string scale, there
may be significant $\alpha'$ corrections.)

As we discuss in section 2, the metrics with positive scalar curvature on
$K$ lie a finite distance away from the moduli space of Ricci flat metrics.
This shows that the negative energy is a nonperturbative effect, which
cannot be seen at any finite order in perturbation theory about the moduli 
space. 
In terms of an effective four dimensional potential, we will see that
(part of) the negative energy density can be described by a scalar
field with a negative potential which falls off exponentially. Negative
potentials are familiar in supersymmetric
AdS compactifications, but to our knowledge, this is the first time
they have been seen in supersymmetric asymptotically flat compactifications.

Given the existence of negative energy density, it is natural to ask if the
total ADM energy can be negative. Recall that
the total energy for any spacetime which asymptotically 
approaches $M_4\times K$ is well-defined. In the construction outlined above, 
the negative energy grows like the volume of the ball. The transition region
includes positive energy, but one might expect this only grows like the area 
so the total energy could be negative. This would imply that
these supersymmetric compactifications are unstable.  However, this
does not occur. There are positive energy theorems 
\cite{Witten:mf,Boucher:1984yx,Dai} 
which ensure that the total energy remains positive. Coleman and 
De Luccia \cite{Coleman:1980aw}
showed long ago that gravity can stabilize a false vacuum. (This was
applied to supersymmetric vacua in  \cite{Weinberg:id,Cvetic:1992st}.)
They considered a scalar field with a potential with two local 
minima that are close together. We have discovered 
an extreme generalization of this phenomenon: gravity
is stabilizing a false vacuum in a theory with a
potential which is unbounded from below. This is a flat space analog of the
familiar situation for supersymmetric AdS vacua 
\cite{Breitenlohner:jf,Gibbons:aq}.

Despite the fact that the total energy remains positive,
the existence of regions of negative energy does have physical
consequences which we begin to explore. These
include new instabilities at finite temperature, and possible violations
of cosmic censorship.

An outline of this paper is as follows. In section 2 we review the mathematical
results about the existence of metrics of positive scalar curvature that
we will need. The next section shows how to use these results to
construct configurations with negative energy density. It also
gives a qualitative
discussion of the effective four dimensional potential that results from 
compactification. To gain intuition for how the total energy can remain
positive given the unbounded negative potential, we discuss a simple
model of four dimensional gravity coupled to a single scalar field
in section 4. The following section reviews the positive energy theorems which
guarantee that the ADM energy is never negative. In section 6 we discuss
some of the physical consequences of the negative energy density. The
final section contains remarks about other applications of our results,
e.g., to the stability of de Sitter spacetime, and using 
S-duality to obtain negative energy density even in K3 compactifications.

\setcounter{equation}{0}
\section{Metrics with positive scalar curvature}

Since metrics with positive scalar curvature play such an important role
in our construction, we review here some of the main mathematical results
that we will need. (For a more complete discussion of results up to 1989,
see \cite{Lawson89}.) Let $K$ be a compact manifold.
At first sight, finding a metric whose scalar
curvature is positive at every point on $K$ sounds easy, since it is only one
scalar inequality on the entire metric. Indeed, every compact $K$ (of
dimension $n>2$) admits metrics
with negative scalar curvature. However it is known that there is a
topological obstruction to the existence of metrics with  positive scalar
curvature. For example, the torus $T^n$ (for any $n$) does
not admit any metric with\footnote{In this section we will only consider
the scalar curvature of $K$ and drop the subscript $K$ on $\R$.} 
$\R>0$ \cite{Schoen79}. The same is true for $K3$.
In this case the proof is easy: 
The index for the Dirac operator on $K3$ is
nonzero. This means that for every metric on $K3$, there are nonzero
spinors which solve the Dirac equation: $\gamma^i \nabla_i \e =0$.
However, if we square the Dirac
operator we get
\beq
\nabla^2 \e - {1\over 4}\R \e =0. 
\eeq
Multiplying by the adjoint spinor and integrating over $K$ yields
\beq
\int |\nabla \e|^2 + {1\over 4}\R |\e|^2 =0. 
\eeq
This shows that the metric cannot have positive scalar curvature. This
argument clearly applies to any manifold for which the index of the Dirac
operator, $\hat A(K)$, is nonzero.

There is a generalization of $\hat A(K)$ which
has been shown to completely characterize when a (simply connected, spin)
manifold admits a metric
of positive scalar curvature. It is usually denoted $\alpha(K)$, and was
first introduced in 1974 by Hitchen \cite{Hitchen74}
who showed that if there is a metric 
with $\R>0$, then $\alpha(K) =0$. The converse was established 
by Stolz in 1990 \cite{Stolz90}:
If $\alpha(K) =0$, there always exists a metric with
$\R>0$.
If the dimension of the manifold is a
multiple of four,  $\alpha(K)$ is proportional to $\hat A(K)$.
In general, the definition of $\alpha(K)$ is more involved, but for
our purposes, its most important property is that it can be nonzero
only in dimensions $n= 0,1,2,4\ {\rm mod}\ 8$. (This assumes $n>4$.)
So there is a possible obstruction to positive scalar curvature
only in these dimensions. Even in these dimensions,
this obstruction only applies
to spin manifolds: If $K$ does not admit spinors, then it always admits a metric
of positive scalar curvature \cite{Lawson80}.
Taken together, these results show the following: 

{\it Every simply connected manifold of dimension 5, 6, or 7 admits
a metric with
positive scalar curvature.}

The situation for nonsimply connected manifolds is much more complicated
and still not well understood \cite{Dwyer02}.
We have already remarked that
the torus does not admit a metric with $\R>0$ even in $5,6,7$ dimensions.
However, it has been shown that if $K$ is simply connected and admits a metric
of positive scalar curvature, then $K/Z_p$ also admits a metric of
positive scalar curvature if $p=2$ or an odd integer \cite{Rosenberg94}.
It is likely that all six dimensional Calabi-Yau spaces (whether simply 
connected or not) admit metrics of positive scalar curvature
\cite{McInnes:2001xp}.

Let $K$ be a compact $n$ dimensional
manifold admitting metrics of positive scalar curvature,
and consider the space
of all smooth Riemannian metrics on $K$. There is the following 
simple description of
this space.
First note that the space of all metrics is connected: Any metric can
be continuously deformed into any other metric. This is most easily seen
by characterizing each metric at a point by an orthonormal frame, and noting
that one set of linearly independent vectors can certainly be continuously 
connected to another. Second,
under a conformal rescaling $\tilde g_{mn} = 
\psi^{4/(n-2)} g_{mn}$ the scalar curvature transforms as
\be\label{rescale}
\tilde \R = \psi^{-4/(n-2)} [\R - a \psi^{-1} \nabla^2 \psi ]
\ee
where $a=4(n-1)/(n-2)$.
Now consider the lowest eigenvalue of the conformally
invariant Laplacian: 
\be
-a\nabla^2 \psi + \R\psi = \lambda_0 \psi. 
\ee
The corresponding eigenfunction is nonzero and can be used as a conformal
factor in (\ref{rescale}). The resulting scalar curvature has the same sign
as the eigenvalue $\l_0$.
Although this
construction does not yield constant scalar curvature, the Yamabe conjecture
and subsequent proof \cite{Lee87} shows that one can always
conformally rescale these metrics to ones with
constant scalar curvature.
This is very useful since $\l_0$ is just a function on the space of metrics
and divides it up into two regions depending on whether the metric can
be rescaled to positive scalar curvature or not. 

By continuity, the boundary of the region with $\R>0$ contains metrics
with $\R=0$. In addition, there
is the possibility of ``islands of $\R=0$" metrics a finite distance away
from the positive scalar curvature metrics (The subspace of zero scalar 
curvature metrics need not be connected).  
However, those metrics must be Ricci flat, $\R_{mn} =0$. To see this, note that
if $\R_{mn} \ne 0$ we can perturb the  metric
by a multiple of the Ricci tensor,
$\delta g_{mn} = b \R_{mn}$. Since $\R_{mn}$ is traceless and divergence
free, the change in the scalar curvature under this perturbation is
$\delta \R = -b \R_{mn} \R^{mn}$. So by changing the sign of the coefficient
$b$, we can construct nearby
metrics with either positive or negative scalar curvature.
This shows that any metric with zero scalar curvature but nonzero Ricci
tensor, must lie on a boundary between positive and negative $\R$ 
metrics.\footnote{This is one way to see that the only metrics on $K3$ with
zero scalar curvature are the Ricci flat metrics.}

\begin{figure}[htb]
     \centerline{\epsfxsize=8.6cm 
       {\epsfysize=6.0cm
             \epsffile{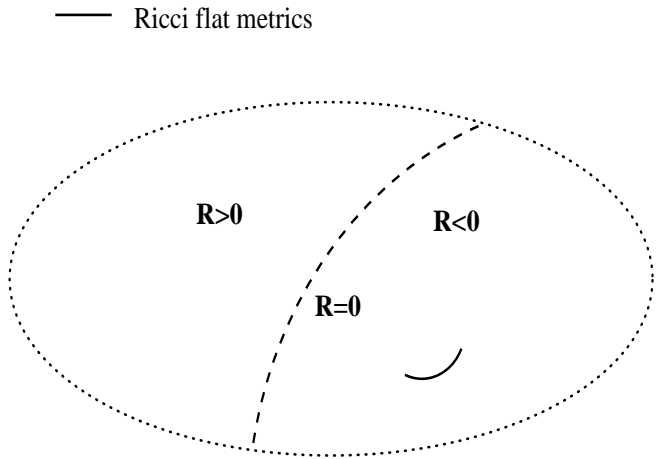}}}
      \caption{The space of metrics for $K$ is shown, where $K$ is  e.g. 
a simply connected Calabi-Yau 
or $G_2$ manifold. The moduli space of Ricci-flat metrics is surrounded by
metrics that can be conformally rescaled to have $\R<0$. }
\label{1}
\end{figure}

A Ricci flat metric could also lie on the boundary of a region of $\R>0$ 
metrics. So a natural question to ask is: Where is the moduli space
of Ricci flat metrics in the usual supersymmetric compactification?
We now show that they must lie a finite distance away from the 
positive scalar curvature metrics (see Fig.~\ref{1}).
Supersymmetry guarantees that there are no tachyons in the perturbative
spectrum. However all Ricci flat metrics which can be perturbed to
both positive and negative scalar curvature must have tachyons.
This can be seen as follows.
 A general perturbation can be divided into a pure
trace, and transverse traceless pieces. A pure trace perturbation produces
a first order change in the scalar curvature proportional to $\nabla^2 h$.
This integrates to zero so it clearly cannot have definite sign.
A transverse traceless perturbation, $h_{mn}$,
does not change $\R$ to first order. To study its second order effects,
consider the functional
\beq
S=\int_K \R \sqrt g.
\eeq
Its first variation is
\beq
\delta S = \int_K (\R_{mn} - {1\over 2} \R g_{mn})h^{mn} \sqrt g. 
\eeq
This vanishes for all perturbations since the background is Ricci flat.
The second variation is 
\beq
\delta^2 S = \int_K h^{mn} \Delta_L h_{mn} \sqrt g
\eeq
where $\Delta_L$ is the Lichnerowicz operator coming from the linearized
Einstein equation. It is now clear that $\Delta_L$ must have both positive
and negative eigenvalues in order for $\delta^2 S$ to have either sign.
But this means that if we start with the product spacetime
$M_4\times K$ with metric $ds^2 = \eta_{\mu\nu} dx^\mu dx^\nu
+ g_{mn}(y) dy^m dy^n $, some metric perturbations of the form
$\phi(x) h_{mn}(y)$ will be tachyonic. 
Thus the moduli space of supersymmetric, Ricci flat metrics must
lie a finite distance\footnote{There probably exist manifolds $K$ that admit 
Ricci flat metrics on the boundary of an $\R>0$ region, but they are necessarily
not supersymmetric.} away from the metrics with $\R>0$, as shown in Figure 1.

\setcounter{equation}{0}
\section{Negative Energy Density in Four Dimensions}

As described in the introduction, the existence of metrics of positive
scalar curvature on $K$ leads to effective negative energy density 
in the four dimensional theory. In this section, we explain this in
more detail and describe some qualitative features of the four dimensional
effective potentials which arise.

We want to consider vacuum solutions which asymptotically approach
$ M_4 \times K$,
where $K$ is, e.g., a simply-connected Calabi-Yau space.
We can characterize these solutions in terms of initial data. To minimize
the mass, we will consider time symmetric initial data which are
spherically symmetric on $\RR$. There are many ways to construct
solutions to the initial value  constraint $\R=0$ which have regions of
negative energy density on $\RR$. We now describe one approach.

Consider the metric
\be \label{ansatz}
ds^2 = \left( 1-{2m(r)\over r} \right)^{-1} dr^2 +r^2 d\Omega_2^2
+g_{mn} (r,y) dy^{m} dy^{n}  
\ee
where the indices $m,n$ label the extra compact dimensions.
The metric $g_{mn} (r,y)$ denotes a one parameter family of metrics on $K$. 
The Ricci scalar of (\ref{ansatz}) is 
\beq \label{rscalar}
{\cal R} & = & 
- \left( 1-\frac{2m(r)}{r} \right)\left[-\frac{1}{4} \partial_{r} g^{mn}
\partial_{r} g_{mn} + g''
+\frac{2}{r} g' +
\frac{1}{4} (g')^2 \right]\nonumber\\
& &  
+ \R_{K}  
+\partial_{r} m \left( \frac{4}{r^2} +\frac{g'}{r}  
\right)
-\frac{m}{r^2} g' 
\eeq
where $g' \equiv g^{mn} \partial_r g_{mn}$, $g'' = \partial_r g'$, and
$\R_{K} $ is the scalar curvature of $g_{mn}(r,y)$ at fixed $r$.

Inside some region $r<R_0$, we choose $g_{mn} (r,y)$ to be independent of
$r$, and equal to some metric with $\R_K=2V_0$, a positive constant.
In this case, the constraint
$\R=0$  reduces to $\partial_r m = -V_0 r^2/2$ which is easily solved
for $m(r)$ yielding a region of constant negative energy density. We now
pick a radius $R_1>R_0$ and choose any path in the space of metrics which 
connects our positive scalar curvature metric $g_{mn}(R_0,y)$ 
to a metric on the moduli space, $g_{mn}(R_1,y)$. In general, we cannot 
solve $\R=0$ for $m(r)$ because there is nontrivial $y$ dependence. 
However, we can
find an $m(r)$ so that $\R \ge 0$. We can either view this as nonvacuum
initial data for string theory, by adding say a dilaton with $\varphi=0$ and
$\dot \varphi^2 = \R$, or we can obtain vacuum initial data by a subsequent
conformal rescaling of 
the nine dimensional metric (\ref{ansatz}). Let $\widetilde
{ds}^2 = e^{4\psi/7} ds^2$, then  the change in the scalar curvature is
given by (\ref{rescale}).
So if $\psi$ is a solution to the conformally invariant
Laplace equation in nine dimensions
\be
-\nabla^2 \psi + {7\over 32} \R\psi = 0
\ee
then the rescaled  scalar curvature vanishes. In order for the rescaled metric
to be nonsingular and asymptotically flat, we need a solution $\psi$
which is nonvanishing and goes to one at infinity. One can show that
such solutions always exist when $\R\ge 0$ \cite{Witt}.
This conformal rescaling can only decrease the total energy since
the ADM mass changes by
\be
\Delta M \propto -\oint \nabla \psi \propto -\int \R\psi <0. 
\ee

 From the four dimensional viewpoint, the metric on $K$ is like an
 (infinite)
 collection
of scalar fields with potential $-\R_K$. 
Qualitatively, this potential has a local minimum at zero when $g_{mn}$
is on the moduli space. There is then a finite positive barrier separating
this minimum 
from a region where the potential is negative. Since the potential is just
the scalar curvature, the height of the barrier is roughly $1/L_K^2$ where 
$L_K$ is a characteristic size of $K$. Thus large Calabi-Yau spaces have small
potential barriers. The width of the potential is harder to estimate
since it depends on mathematical details about the space of metrics on $K$
which are not yet known.
For example, a key open question is: How close does the 
moduli space of Ricci flat metrics come to the region of positive scalar
curvature metrics? The positive energy theorems we discuss later can be
used to give some information about this distance.

Once one reaches a metric of constant scalar curvature $\R_K =2V_0>0$, one can
always rescale the metric by a conformal factor which is
constant on $K$, to increase the
curvature and make the effective three dimensional energy density more
negative. We can easily compute the effective potential for this mode.
Let us start with a product metric on $\RR \times K$, $ds^2 = ds_3^2 +
ds^2_K$. Let $\phi$ be a function depending only on $\RR$. The scalar
curvature of the metric
\be
ds^2 = e^{-n\phi} ds^2_3 + e^{2\phi} ds_K^2
\ee
is
\be
\R = e^{n\phi}[\R_3 +2V_0 e^{-(2+n)\phi} - {n(n+2)\over 2} (\nabla \phi)^2]
\ee
where $n$ is the dimension of $K$.
The second term on the right is just the scalar curvature of the rescaled
metric on $K$.
The vacuum constraint is $\R=0$, and in $3+1$ dimensions, the energy density
is $\rho =\R_3/2$. So we obtain
\be 
\rho = {n(n+2)\over 4} (\nabla \phi)^2 - V_0 e^{-(2+n)\phi}. 
\ee
Rescaling $\phi$ to have a standard kinetic term we get
\be
\rho = {1\over 2}(\nabla \tilde \phi)^2 - V_0 e^{-\alpha \tilde \phi} 
\ee
where 
\be\label{defalpha}
\alpha^2 = {2(n+2)\over n}.
\ee 
So the potential is not only negative, but
falls off exponentially fast. Notice that for more than one extra dimension,
$2<\alpha^2<6$.

There are other ways to achieve large $\R_K$ (and hence large negative
energy density). We mentioned in the last section
that one can always conformally rescale
a metric on $K$ to one with constant scalar curvature whose sign 
depends on the lowest eigenvalue, $\lambda_0$, of the conformal Laplacian.
If $\lambda_0 \le 0$, the rescaled metric is unique for fixed  volume.
In other words, the only remaining freedom is the
constant rescalings discussed above. However, if $\lambda_0 >0$, the
conformally related metric with constant positive curvature is generically
not unique,
even when the volume is fixed.
There is a maximum value of the scalar curvature in each conformal
equivalence class, but it can be arbitrarily big in nearby conformal classes.
More precisely, given a conformal class with $\lambda_0 >0$, for any
$\varepsilon>0$ and any large number $N$ there is another conformal class
within $\varepsilon$  of the first (in a $C^0$ topology) with a metric
of unit volume and constant
scalar curvature larger than $N$ \cite{Pollack}. This surprising result
shows that the maximum scalar curvature in each conformal class is not
a continuous function\footnote{This depends on the topology one puts on the
space of conformal classes. In a stronger topology, this function will probably
be continuous \cite{Pollack2}.}.

There is clearly a lot of freedom in the construction of configurations
with negative energy density. One can choose any path in the space of
metrics from the moduli space to the region with $\R_K>0$,
and then choose any path in the region of positive scalar curvature metrics
to make $\R_K$ large. One approach is to divide the motion up into
nonconformal motion and conformal rescalings. For the nonconformal motion,
there are two natural choices:
 One is to choose a path that only
includes metrics of constant scalar curvature and fixed volume; the other
is to keep a local volume element fixed. This second option greatly simplifies
(\ref{rscalar}) since $g'=0$, but since the scalar curvature does not
remain constant, one needs to rescale the final metric by
a position dependent conformal factor to obtain 
constant $\R_K>0$.

\setcounter{equation}{0}
\section{Scalar Field Model}

In the next section we will review the theorems which show that
despite the unbounded negative energy density,
the total energy must remain positive.
To gain intuition into how this is possible,  we now
consider a simple model of four dimensional gravity coupled to a
single scalar field with potential $V(\phi)$.
To model the situation
we have for the compactifications, we
consider  potentials $V(\phi)$ which are negative in a region of $\phi$
which is separated by a positive barrier from  a local minimum at
$\phi_f >0$ where the potential vanishes.
We consider
configurations where $\phi \rightarrow \phi_f$ asymptotically,
and ask whether the total energy can be negative. In the absence of 
gravity, the answer is clearly yes: Inside a large ball of radius $R_0$,
one can 
\begin{figure}[htb]
     \centerline{\epsfxsize=9.6cm 
       {\epsfysize=7.5cm
             \epsffile{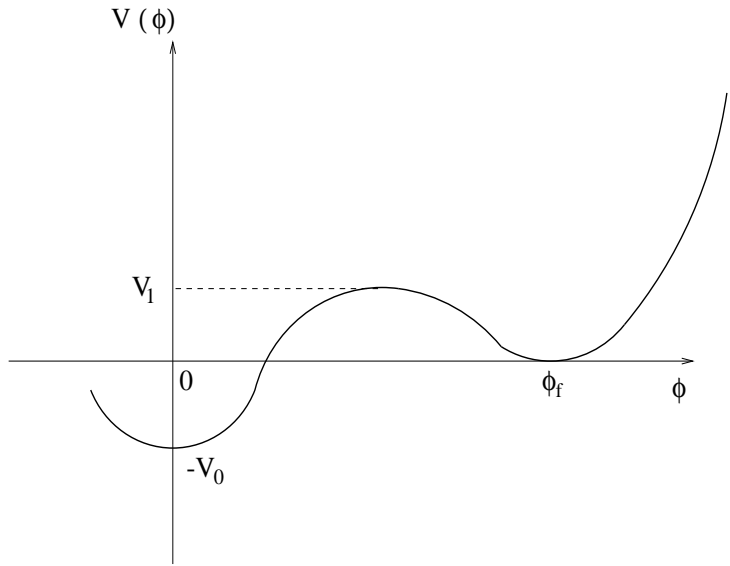}}}
      \caption{A potential with a negative minimum at $\phi=0$ that   
is separated by a positive barrier from a local minimum at $\phi_f$.}
\label{2}
\end{figure}
keep $\phi$ constant at a value where the potential is negative, and
then have $\phi$ approach $\phi_f$ in a transition region. Since the
negative energy in the ball grows like the volume and the positive energy
in the transition region only grows like the area, the total energy
can clearly be negative. However, as first shown by Coleman and De Luccia
\cite{Coleman:1980aw},
gravity can modify this picture.

Since we want to minimize the energy, we set all the time derivatives
to zero. 
For time symmetric initial data the constraint equations reduce to
\be
\ ^{(3)}{\cal R} = 2 \rho
\ee
where
\be
\rho = \frac{1}{2} g^{ij}\phi_{,i}\phi_{,j} +V(\phi). 
\ee
Since spatial gradients raise the energy,
we consider a spherically symmetric configuration with metric
\be
ds^2 = \left(1-{2m(r)\over r}\right)^{-1} dr^2 + r^2 d\Omega^2_2. 
\ee
The constraint then
yields the following equation for the ``mass" $m$ as a function of the radius,
\be \label{mscalar}
m' +\frac{1}{2} mr\phi'^2 = \frac{1}{2} r^2 \left[V(\phi)+\frac{1}{2}\phi'^2
\right]
\ee
where prime denotes the ordinary derivative with respect to $r$.
The ADM mass is simply the value of $m(r)$ at infinity.

We can choose $\phi(r)$ arbitrarily and solve (\ref{mscalar}) for $m(r)$. 
The general solution is
\be\label{gensoln} 
2m(r) =  e^{-\int_{0}^{r} \hat r\phi'^2/2\, d\hat r} 
\int_{0}^{r} e^{\int_{0}^{\tilde r} \hat r\phi'^2/2\, d\hat r} \left[V(\phi) +
{1\over 2}\phi'^2 \right] 
\tilde r^2 d\tilde r.  
\ee
Consider now a potential with a negative 
minimum at $\phi =0$ that is separated by a positive barrier
from a local minimum at $\phi_f $, 
where the potential vanishes.
We call the height of the positive barrier $V_1$, and the depth of the
negative minimum $V(0)=-V_0$ 
(see Fig.~\ref{2}). 
We pick two
radii $R_1>R_0$ and  set
$\phi=0$ for $r<R_0$, $\phi= \phi_f$ for $r>R_1$,
and let $\phi$ be any smooth increasing function for $R_0<r<R_1$.
The solution for $m(r)$ for this configuration is 
\be 
2m(r) = e^{-\int_{R_0}^{r} \hat r\phi'^2/2} \left[-\frac{1}{3} V_0 R_0^3
+\int_{R_0}^{r} \left (V(\phi) +
{1\over 2}\phi'^2 \right ) e^{\int_{R_0}^{\tilde r} \hat r\phi'^2/2} 
\tilde r^2 d\tilde r \right]. 
\ee
One sees that the positive energy in the barrier is enhanced relative
to the negative energy by an exponential factor.
We shall see that the ADM mass can nevertheless be negative provided
that $V_0/V_1$ is sufficiently large. To identify the approximate criteria
for this to be the case, we minimize the exponential factor
by taking 
\be
\phi (r) = \mu \ln (r/R_0)
\ee
in the transition region. This fixes the relation between $R_0$ and $R_1$
to be
$R_1=R_0 e^{\phi_{f}/\mu}$. The ADM mass $M=m(\infty)=m(R_1)$ is given by
\beq
2M &= & e^{- \mu \phi_{f}/2} \left[ -\frac{1}{3}V_0 R_0^3
+\frac{\mu^2 R_0}{( \mu^2 +2)} \left(e^{\phi_{f}(\mu^2 +2)/2\mu}-1 \right)
 \right.\nonumber\\
& & +\left. \frac{R_0^3}{\mu} \int_{0}^{\phi_{f}} e^{x(\mu^2+6)/2\mu}V(x)dx 
\right]. 
\eeq
Notice that the last term, coming from the transition region,
scales like $R_0^3$, just like the first term. 
If the slope $\mu$  and the width of the potential barrier $\phi_f$ are
$ \sim {\cal O}(1)$, then the integral is $ \sim V_1$.
For $R_0 \gg 1$, the condition for the ADM mass $M$ to be negative is 
then simply $V_{0} >V_1$.
For an arbitrary slope $\mu$ and width  $\phi_f$,
and assuming $V(\phi) =V_1$ for 
$0<\phi<\phi_{f}$, we have $M<0$ if (again for $R_0\gg 1$)
\be
\frac{V_0}{V_1} > \frac{6}{(\mu^2 +6)}
\left(e^{\phi_{f}(\mu^2+6)/2\mu}-1 \right). 
\ee
The right hand side is minimized for $\mu$ of order one.

Intuitively, one reason that it is harder to get negative total energy
in theories with gravity is that a constant negative energy
density produces a hyperbolic space with constant negative curvature. In this
space, the volume inside a ball of large radius is equal to 
the volume of a thin shell at this radius with a width equal to the scale of
curvature. 

As we saw in the last section, the potentials which arise
from Calabi-Yau compactifications are unbounded from below.
\begin{figure}[htb]
     \centerline{\epsfxsize=9.6cm 
       {\epsfysize=8.0cm
             \epsffile{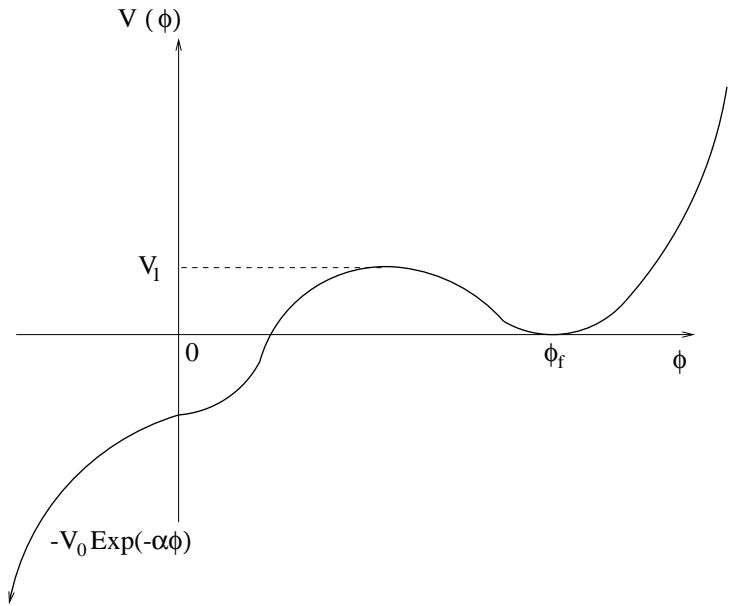}}}
      \caption{A potential which is unbounded from below for 
 negative values of $\phi$.}
\label{3}
\end{figure}
Now suppose we modify the potential for negative $\phi$ so that it
falls off exponentially $V = - V_0 e^{-\alpha\phi}$~(see Fig.~\ref{3}). 
Since this potential is unbounded from below, one might expect
that one can always construct negative energy solutions. But this is not
the case. We now show that there is a critical value,  $\alpha^2 =6$, such
that if $\alpha^2<6$ the energy can stay positive by an order one 
barrier. 

As we saw above, the key point is how low can we make the energy density
inside a ball of radius $R_0$, if we require $\phi(R_0)=0$. If we keep
$\phi=0$ everywhere inside the ball, the energy density would be $-V_0$.
If we include negative values of $\phi$, we can access the growing negative
potential, but we have to pay the price of the derivative contributions
to the energy and the exponential enhancement of the barrier.
Clearly, to make the
energy density as negative as possible, we want $\phi$ to change
slowly. This minimizes the positive $\phi'^2$ contribution to the
energy and increases the region over which the negative potential
contributes. However, the most important effect is the exponential
factor. To minimize the exponential, we take
$\phi= \mu\ln (r/R_0)$. This configuration is singular at the origin, but
we will smooth it out by setting $\phi$ equal to a constant $ -\phi_0 <0$
for $r<r_0$. Thus $\phi_0 = \mu \ln (R_0/r_0)$. Since $\phi$ is constant,
the exponential vanishes for $r<r_0$ and for $r_0 < r< R_0$ we have
\be
 \int_{r_0}^r r\phi'^2 = \mu^2 \ln (r/r_0). 
\ee
Using the general solution (\ref{gensoln}) we get
\beq
2m(R_0)& =& -{ r_0^3V_0\over 3} \({r_0\over R_0}\)^{{\mu^2\over 2} -\alpha\mu}
   -{2V_0\over\mu^2 -2\alpha\mu +6}
   \left [ R_0^3 - r_0^3 \({r_0\over R_0}\)^{{\mu^2\over 2} -\alpha\mu}\right]
   \nonumber\\
 &  &+{\mu^2\over \mu^2 +2}
   \left [ R_0 - r_0 \({r_0\over R_0}\)^{\mu^2/2}\right]. 
\eeq
The first term is the contribution to the energy from the inner region
$r<r_0$, the second term is the contribution from the potential in
the region $r_0 <r<R_0$, and the last term is the contribution from the
derivative terms. To make the energy density inside $R_0$ as negative as
possible, we should choose $r_0 \ll R_0$. Then 
\be\label{negdensity}
{m(R_0)\over  R_0^3} =- {V_0\over \mu^2 - 2\alpha\mu +6}. 
\ee
For $\alpha^2<6$, this is maximized at $\mu=\alpha$, when the right hand
side becomes
$-V_0/(6-\alpha^2)$. If $\alpha^2 \ge 6$, the energy
density inside $R_0$  can be made arbitrarily negative, by choosing $\mu$
so that the denominator in (\ref{negdensity}) is as small as one wants. In 
this case, no finite positive barrier can prevent the total ADM energy from 
being negative. Recall that the values of $\alpha$ that arise from 
compactification (\ref{defalpha}) all satisfy $\alpha^2 <6$.

One can show that $\phi(r) = \alpha \ln (r/R_0)$ is indeed an extremal path
for the potential
energy, given an exponential potential $V = -V_0e^{-\alpha\phi}$.
If one changes this path by a small perturbation $\varepsilon(r)$ with 
$\varepsilon(R_0)=0$,
then the $R_0^3$ contribution to the energy does not change to first order
in $\varepsilon$.

\setcounter{equation}{0}
\section{Positive energy theorems}

Despite the fact that the energy density can be arbitrarily negative,
the total energy must remain positive. This follows from 
a simple extension of Witten's proof \cite{Witten:mf}
of the positive energy theorem\footnote{The proof by Schoen and Yau
\cite{SY} does not extend to ten or eleven dimensions.}. We first review the
original four dimensional version, and then discuss its extension 
to higher dimensions. Consider a 
four dimensional  asymptotically flat
spacetime with matter satisfying the dominant energy
condition: $T_{\mu\nu} t^\mu \tilde t^\nu \ge 0$, where $t^\mu$ and
$\tilde t^\nu$ are any two future directed timelike vectors. Let $\Sigma$
be a nonsingular, asymptotically flat spacelike surface and let $D_i$
be the projection of the spacetime covariant derivative into $\Sigma$.
Let $\e$ be a solution of the spatial Dirac equation $\gamma^i D_i \e=0$
which approaches a constant spinor $\e_0$ at infinity. Then one can
show that 
\be\label{pet}
\oint_\infty \e^\dagger D_i \e\ dS^i = \int_\Sigma [(D_i \e)^\dagger (D^i \e)
+ {1\over 2}T_{\mu\nu} n^\mu (\e^\dagger \gamma^\nu \e)] d\Sigma
\ee
where $n^\mu$ is the unit normal to $\Sigma$. The right
hand side is nonnegative, and the surface integral at infinity is proportional
to $P_\mu t_0^\mu$ where $P_\mu$ is the total
ADM four momentum and $t_0^\mu =\e^\dagger_0 \gamma^\mu \e_0$ is a timelike
(or null) vector. This proves that the total ADM energy cannot be negative.
If the total energy vanishes, $T_{\mu\nu}=0$, $\e$ is covariantly constant,
and the spacetime must be Minkowski space.
The only subtlety in the proof is showing that asymptotically 
constant solutions to $\gamma^i D_i \e=0$ exist. A plausible argument
was given by Witten and  a rigorous proof was given in \cite{Parker:uy}.

This argument easily extends to 
higher dimensional spacetimes which admit spinors and
have a covariantly constant spinor at infinity. The calculation
which leads from  $\gamma^i D_i \e=0$ to (\ref{pet}) continues to hold
in higher dimensions. Since supersymmetric vacuum compactifications $K$ 
always admit covariantly constant spinors,  solutions
which asymptotically approach $M_4\times K$ cannot have negative total energy.
Furthermore, the only solution with zero energy is $M_4\times K$ itself.
Once again, the only
subtlety is the existence of appropriate 
solutions to the spatial Dirac equation.
For Calabi-Yau compactifications, this has recently been shown rigorously
by Dai \cite{Dai}. It seems surprising that this proof of positive total energy
treats K3 and Calabi-Yau compactifications identically, even though we
have seen that the latter have
four dimensional regions of negative energy density and the former
do not. On a time symmetric surface $\Sigma =\RR \times K$ in a vacuum
solution, the proof only uses the
fact that the scalar curvature of $\Sigma$ vanishes, and does not care
whether there are positive scalar curvature metrics on $K$  or not.
In fact, Witten's proof of positive ADM energy was given before it was even
known that such metrics of positive scalar curvature exist.

From a purely 
four dimensional viewpoint,
Calabi-Yau compactifications
include fields with
potentials that are unbounded from below. These certainly do not satisfy
the dominant energy condition, so the usual form of the
(four dimensional)
positive energy
theorem does not apply. It is well known that for
asymptotically anti de Sitter spacetimes, one can have positive energy even
with potentials which are unbounded from below 
\cite{Breitenlohner:jf,Gibbons:aq}.  
It is much less known that 
this is also possible in asymptotically flat spacetimes.
However this was shown almost twenty years ago by 
Boucher \cite{Boucher:1984yx} (see also \cite{Townsend:iu}).
The idea is to modify the
covariant derivative in Witten's spatial
Dirac equation to include an arbitrary function of the fields.
In the simplest case of a single scalar field, one takes
$\hat D_i = D_i + i W(\phi) \gamma_i $. It then turns out that
one can prove positive energy provided the potential for $\phi$
takes the form 
\be\label{superpot}
V = 2 W'^2 - 3W^2. 
\ee
This is similar to the form of the scalar potential in $N=1$ supergravity,
in terms of the superpotential $W$, but one can do this even in 
nonsupersymmetric theories. 
This potential $V$ can be unbounded from below, but if it has a local
minimum at $V=0$, then there will be asymptotically flat solutions and
the positive energy theorem will hold.

We do not know the exact form of the potentials which arise in Calabi-Yau
compactifications far off the moduli space.   
If we start with Type II string theory, we obtain
four dimensional $N=2$ supergravity coupled to matter, and
previous analysis has mostly focused
on the massless moduli fields. The positive energy theorem can be
used to say something about the shape of the potential away from the
moduli space. Consider
 the simple conformal mode that we discussed in section 3 with
exponential potential. Recall that
the potential was $V= - V_0 e^{-\alpha \phi}$, with $\alpha^2 <6$.
This applies inside the region of positive scalar curvature. Moving from
this region to the moduli space corresponds to crossing a potential
barrier of height $\sim 1/L_K^2$, where $L_K$ is the size of the
Calabi-Yau
space.
A positive energy theorem is possible only if the width of the potential
is large enough. If $V_0 \sim 1/L_K^2$, the width must be at least
of order one.

It was suggested in \cite{DeWolfe:1999cp} that one could take any $V(\phi)$ and solve
(\ref{superpot}) for $W(\phi)$, but clearly global solutions only exist for a
restricted class of $V(\phi)$. We saw in section 4 that 
the exponential potential
$V= - V_0 e^{-\alpha \phi}$ with $\alpha^2 \ge 6$ always has negative
energy solutions and hence a solution for $W$ cannot exist. In contrast,
 for $\alpha^2 < 6$, one can obtain $V$ from 
 $W= W_0 e^{-\alpha\phi/2}$ with $W_0^2 = 2 V_0/(6-\alpha^2)$.
Note that 
the critical value of the exponent which arose in constructing negative
energy solutions is precisely the one which allows  a global
solution for $W$.

\section{Physical consequences of negative energy density}

Even though the total energy stays positive, the existence of
unbounded negative energy density may have serious consequences.
In this section we discuss some of these.

\subsection{Cosmic Censorship}

The existence of negative energy density may lead to violations of
cosmic censorship. That is, one might have nonsingular initial data,
which evolves to singularities that are not hidden inside event
horizons. As an example, consider a theory with a 
single scalar field and negative
exponential potential  $V(\phi) = - e^{\alpha\phi}$ for $\phi>0$
and a local minimum at $\phi_f<0$ where $V=0$. (This is similar to the
example in section 4 except that the sign of $\phi$ has been changed to
remove some unnecessary minus signs.)
Consider time symmetric initial data consisting
of a constant scalar field $\phi= \phi_0 >0$ for $r<R_0$ and then
$\phi$ changes continuously to the local minimum of the potential at $V=0$
at larger radius. Inside the sphere $r=R_0$, the energy density is
a negative constant, so the three geometry has constant negative curvature.
Thus the evolution inside the domain of dependence of the ball of radius
$R_0$ looks like a $k=-1$ Robertson-Walker universe. 
(It is NOT anti de Sitter space, since the scalar field evolves.)
Under evolution, the scalar field rolls down the potential and the
spacetime becomes singular in finite time.
Since this singularity is 
not the result of the usual gravitational collapse, one may not form trapped
surfaces or an event horizon. This would lead to a violation of
cosmic censorship.

To examine this possibility further, let us first estimate the time to the
singularity. Since the metric takes  the
standard Robertson-Walker form, 
$ds^2 = -dt^2 + a^2(t) d\sigma^2$ where $d\sigma^2$ is the metric on
the unit hyperboloid,   the field equations are 
\be\label{fieldeq}
{\ddot a\over a}= {1\over 3}[V(\phi) - \dot\phi^2]
\ee
\be
\ddot{\phi} +{3 \dot a\over a} \dot\phi - \alpha e^{\alpha\phi} =0. 
\ee
We start with $\phi=\phi_0$, and set $\dot\phi=0, \dot a=0$ since
we want time symmetric initial data.  The right hand side
of (\ref{fieldeq}) is negative, so $a$ decreases under evolution and the 
singularity occurs when $a=0$. Since $\phi$ increases 
as it rolls down the potential, the right hand side of (\ref{fieldeq})
is always less than its initial value. Thus we can get an upper bound
on the time to the singularity by solving the simple equation
$\ddot a + a e^{\alpha \phi_0}/3  =0$.
This yields 
\be 
T < \frac{\sqrt 3\pi}{2} e^{-\alpha\phi_0/2}. 
\ee
\begin{figure}[htb]
     \centerline{\epsfxsize=9.6cm 
       {\epsfysize=5.5cm
             \epsffile{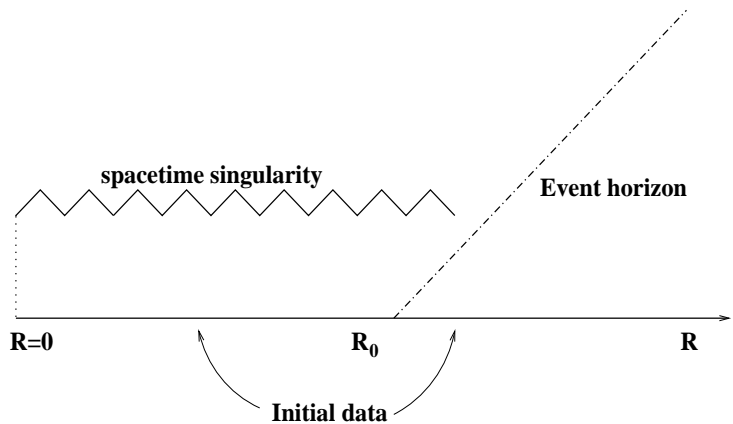}}}
      \caption{If an event horizon encloses the singularity, it must have
      an initial size of order $R_0$.}
\label{4}
\end{figure}

If we choose $R_0 \gg T$, then the singularity will form very quickly
compared to the size of the region. If the singularity lies inside
a black hole, then we can trace the null geodesic generators of the event 
 horizon back to the initial surface where it will form a sphere of radius
approximately $R_0$ (see Fig.~\ref{4}).
Now suppose the total mass is smaller than a Schwarzschild
black hole of size $R_0$. 
The area theorem for black holes only requires the weak energy condition
and hence still holds even in theories with $V(\phi)<0$.
Since the area of the event horizon
cannot decrease during evolution and the mass cannot increase,
when the spacetime settles down there is  not enough mass to support
a black hole large enough to enclose the singularity.
 Inside the domain of
dependence of the initial ball of radius $R_0$,
the singularity will be spacelike like a big crunch. The singularity
is likely to extend outside the domain
of dependence.
There are two
possibilities. Either
the singularity ends, and the endpoint is a naked singularity, or the 
singularity continues all the way out to null infinity cutting off all
spacetime. This is a disaster much worse than naked singularities.

Support for this line of thinking comes from the fact that cosmic 
censorship is violated in certain four dimensional field theories of 
this type \cite{Hertog03}.
We do not yet know if this is realized in Calabi-Yau compactifications, but
it cannot be ruled out.
In the scalar field example, one can show from the field equations that
$\phi$ diverges at the singularity. In the Calabi-Yau context, this would
correspond
to $K$ shrinking to zero size. If one tries to do this keeping the
metric on $K$ Ricci flat, one is likely to form black holes 
\cite{Geddes:2001ht}. Since our data are time symmetric, there will be
singularities in the past as well as the future. If naked singularities
form in this case, it would be interesting to see if they also
form in situations without singularities in the past.

Violations of cosmic censorship could be viewed as a desirable feature
of Calabi-Yau compactifications. If singularities can be visible, one
has the possibility of directly observing effects of quantum gravity.
Of course, the key question is how frequently do naked singularities form.
If they occur too often in certain compactifications, 
then those would be observationally 
ruled out\footnote{It is not difficult to see that cosmic censorship can be 
violated in nonsupersymmetric compactifications, $M_4\times K$, 
if $K$ is a manifold that admits a Ricci flat metric on or close to the
boundary of an $\R_K>0$ region. Of course, such theories are physically 
unacceptable anyway since they do not admit a positive energy theorem.}.

\subsection{New thermal instability}

Since there is only a finite barrier separating the 
supersymmetric vacuum $M_4\times K$
from the region of unbounded negative energy density, one can
ask whether ordinary finite energy processes can cause an instability.
It is unlikely that a two particle collision at high energy can trigger
this instability. This is because the classical process of the field going
over the barrier is described quantum mechanically by a coherent state
which involves many quanta. However, it is possible to see this
instability in a thermal state at nonzero temperature.

At nonzero temperature, there is always an instability for nucleating a 
black hole \cite{Gross:cv}. There is a gravitational instanton describing this
process semiclassically. The instanton is the euclidean Schwarzschild
black hole with the prescribed temperature, and the nucleation rate is given
by the action of this instanton.
This yields $\Gamma \sim e^{-M^2} = e^{-1/T^2}$, so the nucleation
of black holes is highly suppressed at temperatures much below the Planck 
scale ($T_{pl} \sim 10^{32} K$).

The existence of negative energy density in Calabi-Yau compactifications
means there is a new instability corresponding to thermal fluctuations causing
the field to jump to the top of the barrier in some region of space 
and rolling over\footnote{If the width of the potential is small enough, this
instability might manifest itself in terms of quantum tunneling through
the barrier.}. The height of the barrier is
roughly $1/L_K^2$ where $L_K$ is the size of the Calabi-Yau space.
If only a small region has a thermal excitation of
this amount, it just disperses. But if a region of size $L_K$ undergoes
such a thermal excitation it will become unstable. This is 
because this region will now act like a horizon sized patch of de Sitter.
Typically, the region briefly expands before the field rolls down the 
potential. If the field rolls down on the other side,
the region will collapse down to a singularity
\cite{Banks:1984cw,Felder:2002jk}. 
The energy in the original thermal excitation is 
roughly $E \sim L_K$, so the decay rate is 
\be
\Gamma \sim e^{-L_K/T}.
\ee 
For low temperatures $T<1/L_K$, this is larger than the rate of black hole
production.  One could try to make this estimate more
precise by looking for a new thermal instanton solution. 
In four dimensional gravity coupled to a scalar field with a potential
as in figure 3, the decay of the false vacuum at finite temperature
would be described by an $O(3)$-symmetric gravitational instanton
that is periodic in Euclidean time. In our case this would be a
ten dimensional euclidean vacuum solution which asymptotically approaches
$\RR \times S^1 \times K$. Whether or not this new thermal instability
is significant in the early universe 
depends on the size of the internal space.

\section{Discussion}
We have reexamined the four dimensional theories resulting from supersymmetric
compactifications of string theory (or just supergravity). Most previous
discussions have focused on the moduli space of vacua
and the associated massless fields. We have shown that if one moves off the
moduli space, in many cases the four dimensional effective potentials are
unbounded below. This applies to all simply connected Calabi-Yau and $G_2$
manifolds, as well as some nonsimply connected ones.

Naively, this suggests that there should be solutions with negative
total ADM energy, but this is not the case. One can view the fact
that the energy of all nonsingular solutions  remains positive as
a generalization of the phenomenon that gravity
can stabilize a false vacuum \cite{Coleman:1980aw}.

We have begun an investigation of the physical effects of the negative
energy density, and have discussed possible violations of cosmic censorship
and new thermal instabilities. In Calabi-Yau compactifications, there
is another type of possible violation of cosmic censorship 
In the moduli space of Ricci flat metrics, one can deform
a smooth metric into a singular one, e.g., at a conifold point. Since there
are solutions of the form $M_4\times K$ for every point on the moduli space,
one may be able to  change the metric on $K$ slowly enough
so that no event horizon forms in the four noncompact directions even though
a singularity develops on $K$. 
The possible violation of cosmic censorship discussed in the previous section 
is quite
different. In our case, a singularity is forming in the four noncompact
dimensions, not just on $K$. Furthermore, while it has been shown that
the conifold singularity is harmless in string theory, the singularities
we have been discussing may not be. However, it should be kept in mind that
in neither case have violations of cosmic censorship been firmly established
for Calabi-Yau compactifications.

There may be other physical consequences of the negative energy density.
If these consequences are too severe, they may show that all these
compactifications are unphysical. A longstanding problem in string
theory is the large number of apparently consistent vacua in the theory.
Conceivably, the negative energy density could cut down this number dramatically.

Although we have focused on asymptotically flat solutions, the existence
of potentials which are unbounded from below is likely to be important
in other contexts as well\footnote{These applications were suggested
by E. Silverstein and L. Susskind.}. 
For example, the recent construction of de Sitter
space as a solution to string theory  \cite{Kachru:2003aw}
involves a Calabi-Yau internal space. Since
space is now compact, there is no positive energy theorem. The negative energy
density we have found
provides a new type of instability of this de Sitter solution.
As another example,
the existence of negative energy regions
surrounded by positive energy in such a way that the total energy remains
positive is reminiscent of quantum field theory \cite{Ford:1999qv}.
In the context of AdS/CFT
one might wonder whether negative energy density in the (quantum) CFT is 
related to negative energy density in the (classical)  AdS supergravity.

Just because $K3$ does not have metrics of positive scalar curvature does
not guarantee that it is free of potential instabilities. 
S-duality relates Type II string theory on certain Calabi-Yau manifolds
to heterotic string theory on $K3\times T^2$ 
\cite{Kachru:1995wm,Ferrara:1995yx}. In some of these cases, it is known
that the Calabi-Yau space
admits a metric with $\R_K>0$. The fact that the Type II
string has regions of negative energy density means that the same thing must
be true for the heterotic string on $K3\times T^2$. Of course the negative
energy density will not arise in the same way, and  may be a
strong coupling effect.

Another reason for questioning the stability of $K3$ (as well as Calabi-Yau)
compactifications is some recent work of Penrose \cite{Penrose02}. He
argues that all curved compactifications should  have
the property that generic finite perturbations will produce  curvature
singularities. This is worthy of further investigation.

\bigskip

\centerline{{\bf Acknowledgments}}
It is a pleasure to thank M. Cvetic, X. Dai, J. Isenberg, S. Rey,
R. Roiban, J. Polchinski, I. Singer, R. Schoen, K. Schleich, K. Skenderis,
M. Srednicki,
J. Walcher, N. Warner, G. Wei, D. Witt, H. Verlinde, and S.T. Yau for 
discussions. We also thank R. Penrose for stimulating our interest in
the possible instability of compact extra dimensions.
This work was supported in part by NSF grant PHY-0070895 and 
a Yukawa fellowship.

\end{document}